\documentclass[cpp,a4paper,fleqn]{w-art}
\usepackage{times,cite,w-thm}
\usepackage[]{graphicx}
\usepackage{amsmath}
\usepackage{amssymb}
\usepackage{bm}

\newcommand{\beq}{\begin{equation}}
\newcommand{\eeq}{\end{equation}}
\newcommand{\bea}{\begin{eqnarray}}
\newcommand{\eea}{\end{eqnarray}}
\newcommand{\req}[1]{Eq.~(\ref{#1})}
\newcommand{\dd}{\mathrm{d}} 
\newcommand{\gcc}{\mbox{g cm$^{-3}$}}
\newcommand{\hq}{h_\mathrm{q}}
\newcommand{\kTF}{k_\mathrm{TF}}
\newcommand{\nion}{n_\mathrm{ion}}
\newcommand{\nucol}{\nu_\mathrm{coll}}

\begin{document}
\def\copyrightyear{2013}
\DOIsuffix{ctpp.201200094}
\Volume{53}
\Issue{4--5}
\Month{May}
\Year{2013}
\pagespan{397}{405}
\Receiveddate{30 October 2012}
\Reviseddate{21 December 2012}
\Accepteddate{21 December 2012}
\Dateposted{13 May 2013}
\keywords{Dense matter, stellar nucleosynthesis.}


\title[]{Electron screening effect
on stellar thermonuclear fusion}


\author[A. Y. Potekhin]{Alexander Y. Potekhin\footnote{Corresponding
     author: e-mail: {\sf palex@astro.ioffe.ru}}\inst{1,2}}
\author[G. Chabrier]{Gilles Chabrier\inst{2,3}}
\address[\inst{1}]{Ioffe Physical-Technical Institute,
Politekhnicheskaya 26, 194021 St.~Petersburg, Russia}
\address[\inst{2}]{Ecole Normale Sup\'{e}rieure de Lyon,
69364 Lyon Cedex 07, France}
\address[\inst{3}]{School of Physics, University of Exeter,
Exeter, UK EX4 4QL}

\begin{abstract}
We study the impact of plasma correlation effects on
nonresonant thermonuclear reactions for various stellar objects, namely
in the liquid envelopes of neutron stars, and the interiors of white dwarfs, low-mass
stars, and substellar objects.
We examine in particular the effect of electron screening on the
enhancement of thermonuclear reactions in dense plasmas within and beyond
the linear mixing rule approximation as well as the corrections due to quantum
effects at high density.
In addition, we examine some recent unconventional
theoretical results on
stellar thermonuclear fusions
and show that these scenarios do not apply to stellar conditions.
\end{abstract}

\maketitle

\section{Introduction} Thermonuclear reactions play a
crucial role in stellar evolution. Nuclear fusion
rates in stellar interiors can be
significantly enhanced over the binary Gamow \cite{Gamow28}
rates because of the many-body screening effect in the dense
plasma (first recognized by E.~Schatzman \cite{Schatzman48};
for reviews, see \cite{YaSha,Ichimaru93}). 

In the envelopes of neutron stars (NSs) and interiors of
white dwarfs (WDs), where the electrons are strongly
degenerate, the screening effect is usually treated under
the assumption that the electron gas can be considered as a
uniform ``rigid'' background, and the screening is provided
solely by ions. On the other hand, in ordinary stars this
effect is often treated with Salpeter formula, which implies
Debye screening (see, e.g., Ref.~\cite{Bahcall-ea02} and
references therein). The latter approximation is applicable
in a gaseous phase. In the present article, we consider the
electron screening effect on nuclear fusion at arbitrary
electron degeneracy and arbitrary Coulomb coupling of ions
in gaseous and liquid plasmas.

The influence of the electron polarization on the enhancement of
nuclear reaction rates has been studied in some detail
in several papers 
\cite{Salpeter54,ItohTI77,YaSha,SahrlingChabrier98,Kitamura00}.
At the time of those
studies, uncertainties in the reaction rates due to other
factors, viz.\ quantum effects and deviations from the linear
mixing rule in strongly coupled plasmas, as well as
theoretical uncertainties in the nuclear effective potentials at
short distances, were more important than the 
electron-screening
effects. For this reason,
more recent works were aimed at reducing these
uncertainties and mostly neglected the electron
polarization (e.g., 
\cite{Yakovlev-ea06,ChugunovDW09a,ChugunovDW09b}). 
In this paper, we show, however, that the effect of the electron-polarization on
the enhancement factor of the nuclear reaction rates is
typically of the same order of magnitude as the other
recently proposed corrections.

In Sect.~\ref{sect:enh} we compare
different approximations for the enhancement factors.
In Sect.~\ref{sect:res} we describe the results of the
application of the electron-screening correction to the
nuclear reaction rates in
stellar conditions. In Sect.~\ref{sect:deviant} we discuss
the origin of discrepancies between our results and some
other results published recently. Sect.~\ref{sect:concl}
is devoted to the conclusion.

\section{Theory}
\label{sect:enh}

A review of the theory of nuclear fusion in stars
with extensive bibliography was given in
the Nobel lecture by Fowler~\cite{Fowler84}. One
should discriminate between the reactions related to nuclear
resonances and nonresonant reactions. We consider only the
latter ones.
It is customary to write the cross section of binary
nuclear fusion reactions in the form
\beq
   \sigma(E) = {\mathrm{e}^{-2\pi \eta}}\,S(E)/E,
\eeq
where $E$ is the center-of-mass energy of the reacting
nuclei ``1'' and ``2'', 
\beq
   \eta=\sqrt{{E_R}/{E}},
\qquad
   E_R = {(Z_1 Z_2 e^2)^2\,m_{12}}/{2\hbar^2},
\eeq
$Z_j e$ is the charge of nucleus ``$j$'', $e$ is the elementary
charge, $m_{12} = m_1 m_2/(m_1+m_2)$ is the reduced mass, and
$S(E)$ is a function called ``astrophysical factor.''
Then the reaction rate (the
number of fusion events per unit time in unit volume) in the
absence of plasma screening is given by
\beq
   R_{12}=w_{12}\,n_1 n_2\,
    \sqrt{\frac{2}{m_{12}}}
    \int_0^\infty \mathrm{e}^{-2\pi\eta}
       S(E)\,w(E)\,\dd E
  \bigg/
    \int_0^\infty w(E)\,\sqrt{E}\,\dd E,
\label{R}
\eeq
where $n_j$ is the number density of the ions of type ``$j$'',
$w(E)$ is the
statistical distribution function of the center-of-mass
energies of the reacting nuclei, and the factor $w_{12}$ accounts
for statistics: $w_{12}=\frac12$, if nuclei ``1'' and ``2'' are
identical; otherwise $w_{12}=1$. With Boltzmann
statistics, $w(E)=w_\mathrm{B}(E)\equiv T^{-1}\exp(-E/T)$, 
where $T$ is temperature in energy units.

In order to take the plasma screening into account,
it is
convenient to write the radial pair-distribution function for
ions in the form \cite{DeWittGC73}
\beq
   g_{12}(r) = \exp\big( - {Z_1 Z_2 e^2}/{rT} \big)
      \, \exp\big[{H_{12}(r)}/{T} \big] \, ,
\eeq
where the first factor is the Boltzmann formula for an ideal gas,
while the second one shows how the probability of separation of
two chosen ions is affected by the surrounding plasma particles.

It is convenient to introduce parameters
\beq
   \Gamma_{12}={Z_1 Z_2 e^2}/{a_{12}T},
\qquad
   a_{12}={(a_1+a_2)}/{2},
\qquad
   \tau = 3(\pi^2E_R/T)^{1/3} ,
\qquad
   \zeta = 3\Gamma_{12}/\tau .
\label{Gamma12}
\eeq
where $a_j=(3Z_j/4\pi n_e)^{1/3}$ are the ion-sphere radii,
and $n_e$ is the electron number density. As shown in
Ref.~\cite{Ichimaru93}, under the condition $\zeta\ll1$ the
function $H_{12}(r)$ slowly varies on the scale of the classical
turning point distance and
the nuclei behave as classical 
particles. Then the
reaction rate with allowance for the plasma screening is
approximately given by $R_{12}\exp(h)$, where
$h=H_{12}(0)/T$ and $R_{12}$ is expressed by \req{R}
\cite{Salpeter54}. Furthermore, one can prove
\cite{DeWittGC73,Jancovici77,RosenfeldChabrier89,IchimaruKitamura96} that 
$H_{12}(0)$ equals the difference between the excess free
energies $F_\mathrm{ex}$ before and after an individual act of fusion.
 Here, $F_\mathrm{ex} = F - F_\mathrm{id}$, $F$ is the total
Helmholtz free energy, and
$F_\mathrm{id}$ is the free energy of the ensemble of
noninteracting ions and electrons.
In the thermodynamic limit this gives the relation
\beq
   h = \bigg( \frac{\partial}{\partial n_1} 
     + \frac{\partial}{\partial n_2} 
     - \frac{\partial}{\partial n_3} \bigg)
     \,\big[ \nion f_\mathrm{ex}(\{ n_j \}, n_e ,T) \big],
\label{h0gen}
\eeq
where $\nion=\sum_j n_j$ is the total number density of ions,
including number density $n_3$ of composite nuclei, which have
charge number $Z_3=Z_1+Z_2$ and mass  $m_3\approx m_1+m_2$, and
$f_\mathrm{ex}\equiv F_\mathrm{ex}/\nion VT$ is the normalized
excess energy.

In the linear-mixing approximation,
\beq
  f_\mathrm{ex}\approx  f_\mathrm{lm}(\{ n_j \}, n_e ,T)
          \equiv \sum_j x_j f_j(n_e,T).
\label{LMR}
\eeq
Here, $x_j\equiv n_j/\nion$ denotes the number fractions, and
$f_j(n_e, T)$ is 
$f_\mathrm{ex}$ for a plasma containing only the $j$th type of
ions. In this approximation, the enhancement exponent $h$
becomes
\beq
   h_\mathrm{lm} = f_1(n_e, T) + f_2(n_e, T) - f_3(n_e, T).
\label{hLMR}
\eeq
In the model of a rigid electron background, this reduces to
\beq
   h_\mathrm{lm,ii} = f_\mathrm{ii}(\Gamma_1) 
     + f_\mathrm{ii}(\Gamma_2) - f_\mathrm{ii}(\Gamma_3),
\label{hLMRii}
\eeq
where $f_\mathrm{ii}(\Gamma)$ is the normalized excess
free energy of the one-component plasma
and $\Gamma_j=(Z_j e)^2/a_j T$ are coupling parameters of
individual ion species.
In the ion sphere approximation, 
$f_\mathrm{ii}(\Gamma)=-0.9\,\Gamma$, and then $h_\mathrm{lm,ii}$ 
becomes \cite{Salpeter54}
\beq
   h_\mathrm{S} = 0.9\,(\Gamma_3-\Gamma_1-\Gamma_2).
\label{hS}
\eeq

The linear mixing rule works in strongly coupled Coulomb
plasmas, i.e., at $\Gamma_j\gg1$
\cite{HansenVieillefosse76,ChabrierAshcroft90}.
In the opposite limit $\Gamma_j\ll1$ ($\forall j$), 
the Debye-H\"uckel approximation is applicable:
$F_\mathrm{DH} = - VT/12\pi D^3$,
where $D$ is the screening length:
\beq
F_\mathrm{ex}\approx -\frac{VT}{12\pi D^3},
\quad
D^{-2} = \kTF^2 + D_\mathrm{ion}^{-2},
\qquad
D_\mathrm{ion}^{-2} = \frac{4\pi e^2}{T}\,\sum_j n_j Z_j^2,
\qquad
\kTF^2 = 4\pi e^2\, \frac{\partial n_e }{ \partial \mu_e},
\label{DH}
\eeq
where $\mu_e$ is the chemical
potential of the electron Fermi gas.
Using Eq.~(24) of Ref.~\cite{CP98}, one can write $\kTF(n_e,T)$ in
an analytic form. In the two limiting approximations of
nondegenerate electrons
($\kTF^2 \to {4\pi e^2}n_e/T$) and
rigid background ($\kTF\to0$), Eqs.~(\ref{h0gen}) and
(\ref{DH}) give
the Salpeter formula
\cite{Salpeter54}
\beq
   h_\mathrm{DH} = {Z_1 Z_2 e^2}/{DT}.
\label{hDH}
\eeq
Salpeter and Van Horn \cite{SalpeterVH69} proposed a
simple
interpolation between the Debye-H\"uckel and strong-coupling
limits:
\beq
   h_\mathrm{SVH} = \frac{ h_\mathrm{S}\,h_\mathrm{DH}
      }{ 
      \sqrt{h_\mathrm{S}^2 + h_\mathrm{DH}^2}},
\label{SVH}
\eeq
where $h_\mathrm{S}$ and $h_\mathrm{DH}$ are
given by Eqs.~(\ref{hS}) and (\ref{hDH}), respectively.
A more elaborated approximation for the enhancement factor
between the Debye-H\"uckel and strong-coupling limits was
constructed for the rigid
background model in Ref.~\cite{ChugunovDW09b}.

These analytic approximations can be compared to the 
accurate result. We write the normalized excess
free energy in the form $f_\mathrm{ex} = f_\mathrm{lm} +
f_\mathrm{mix}$, where $f_\mathrm{lm}$ is given by \req{LMR}, and
$f_\mathrm{mix}$ is the correction to the
linear-mixing. Then \req{h0gen} gives
\beq
   h_0 = h_\mathrm{lm}
    + \frac{\dd
     f_\mathrm{mix}(x_1+\xi,x_2+\xi,x_3-\xi)
      }{
      \dd\xi} \bigg|_{\xi=0}\,\,,
\label{hmix}
\eeq
where $h_\mathrm{lm}$ is given by \req{hLMR}. The right-hand
side of \req{hmix}
can be written in an analytic form using our fitting
formulae for $f_\mathrm{ex}(n_e,T)$ and
$f_\mathrm{mix}(\{x_j\},\{Z_j\};n_e,T)$ (see  \cite{PC10}
and references therein).

Equation (\ref{hmix}) is obtained assuming that
$H_{12}(r)\approx H_{12}(0)$, which is true for small values
of the parameter $\zeta$ defined in \req{Gamma12}.
When this condition is not satisfied, the classical 
enhancement exponent $h_0$ should be corrected for the
quantum effects. We denote this corrected value $\hq$.
Alastuey and Jancovici \cite{AlaJanco78}
showed that $\hq<h_0$ and developed a perturbation expansion
of $\hq$ in powers of $\zeta$. More recently, Militzer and Pollock
\cite{PollockMilitzer04,MilitzerPollock05} performed
simulations of the contact probabilities in the quantum
regime and extended numerical
results beyond the applicability range of the perturbation
theory \cite{AlaJanco78}. 
Chugunov and DeWitt \cite{ChugunovDW09a} found
that the quantum
effects can be described in the linear-mixing,
rigid-background
approximation by substitution of $\tilde\Gamma_j =
\Gamma_j/t_{12}$ instead of $\Gamma_j$ into
\req{hLMRii}, where
$t_{12}=\left[1+c_1\zeta
    + c_2\zeta^2
     + c_3\zeta^3 \right]^{1/3}$,
$
  c_1 = 0.013\,z^2$,
$
  c_2 = 0.406\,z^{0.14}$,
$
  c_3 = 0.062\,z^{0.19}+1.8/\Gamma_{12}$,
and
  $z =4Z_1 Z_2 / (Z_1 + Z_2)^2$.
An analogous correction is not known for the polarizable
background.
Fortunately, the quantum effects are important only in the domain
of high densities and relatively low temperatures, whereas the
deviations from the linear-mixing and rigid-background
approximations are most important in the opposite case.
Therefore, in order to take all these
effects into account, we multiply the classical
expression (\ref{hmix}) by factor
$q=\tilde{h}_\mathrm{lm,ii}/h_\mathrm{lm,ii}$, where
$h_\mathrm{lm,ii}$ 
and $\tilde{h}_\mathrm{lm,ii}$ are given by \req{hLMRii}
with $\Gamma_j$ and $\tilde{\Gamma}_j$, respectively.

\section{Results}
\label{sect:res}

\begin{figure}
\begin{minipage}[t]{72mm}
\includegraphics[width=\linewidth]{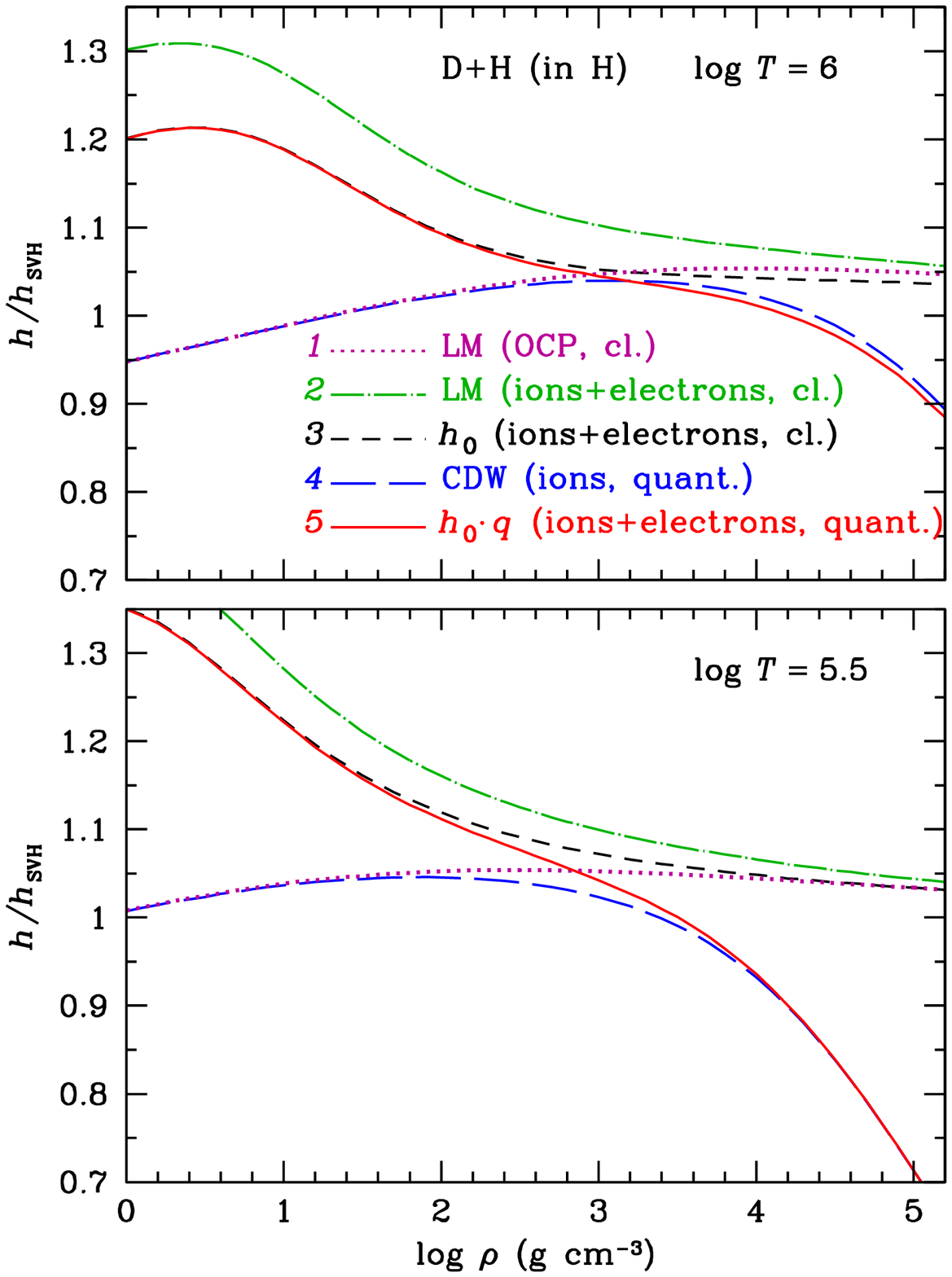}
\caption{Plasma enhancement exponents for deuterium
fusion in hydrogen medium as a function of mass density at
$T=10^6$~K (upper panel) and $10^{5.5}$~K (lower panel) in
different approximations. Three lines show
results for classical nuclei: (\textit{1}) $h_0$ in the linear-mixing, rigid
background approximation [\req{hLMRii}] (dotted lines),
(\textit{2}) $h_0$ for the linear mixing with polarizable electron background
[\req{hLMR}] (dot-dashed lines),
(\textit{3}) $h_0$ beyond the linear-mixing 
approximation with a polarizable electron background
[\req{hmix}], the most general classical
approximation] (short dashes). The other two lines take into
account quantum corrections: 
(\textit{4}) the fit of
Ref.~\cite{ChugunovDW09a} for $\hq$ (long dashes), and
(\textit{5}) the
approximation $\hq = q h_0$ (with $q$ defined in the text),
which includes both the ionic and electronic screening
contributions and takes both the quantum effects 
and the deviations from the linear-mixing rule into account
(solid lines).}
\label{fig:enhqHD}
\end{minipage}
\hfil
\begin{minipage}[t]{73.5mm}
\includegraphics[width=\linewidth]{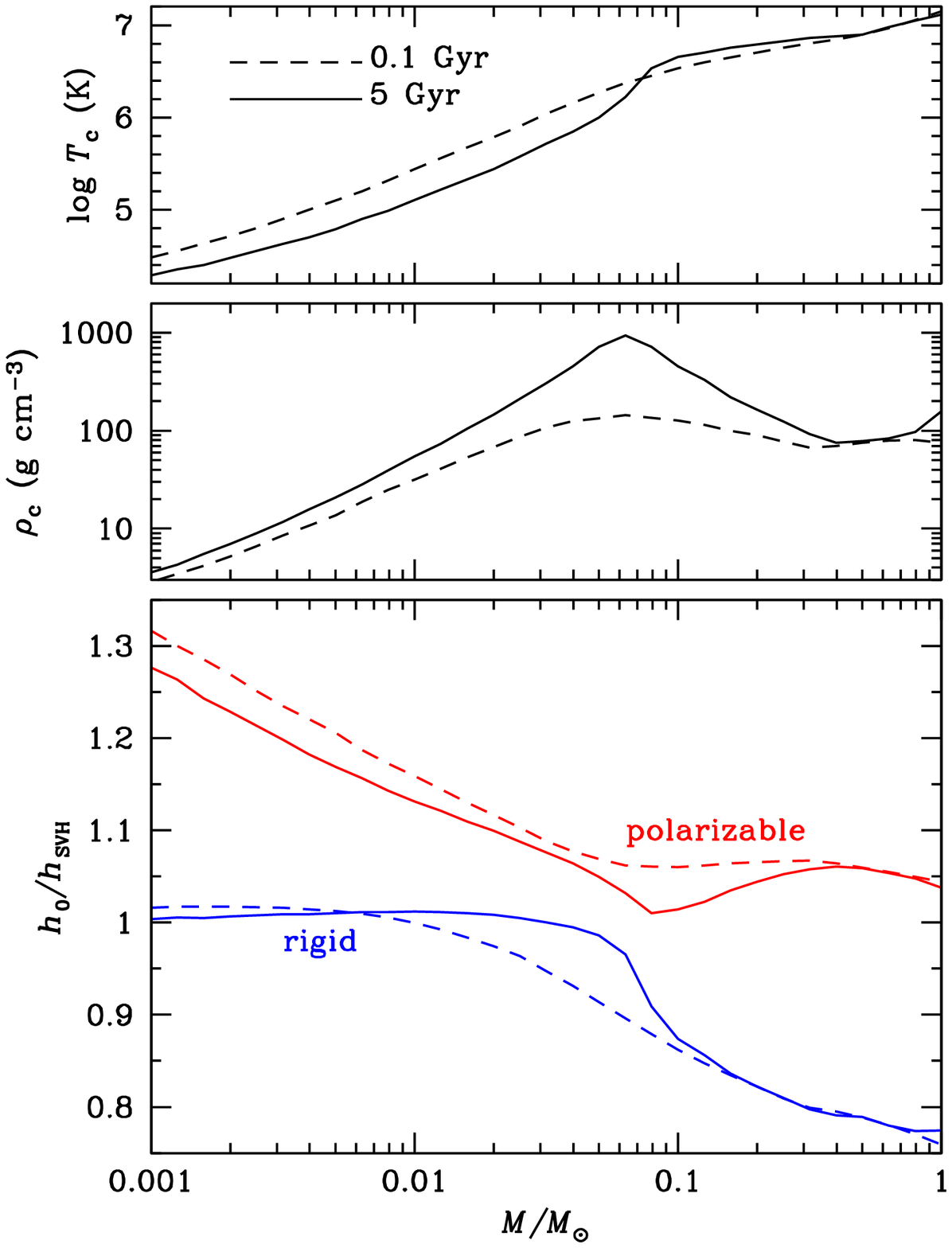}
\caption{Physical conditions (at two ages) and enhancement
exponents (in two approximations) at
the centers of LMSs and SSOs with the solar abundance of heavy
elements. In all three panels solid and dashed lines are drawn 
at the ages 5 Gyr and 100 Myr, respectively.
The top and middle panels reproduce,
respectively, the
central temperature and density along the LMS-SSO mass
range.
The bottom panel shows the normalized enhancement exponent
$h/h_\mathrm{SVH}$ for the deuterium burning
along these temperatures and densities
with (upper pair of curves) and without (lower pair of
curves) account of electron polarization.}
\label{fig:enhLMS1}
\end{minipage}
\end{figure}

\subsection{Degenerate stars}

The electron-screening effects on the enhancement factors and
ignition curves for carbon and oxygen fusion reactions
in the liquid layers of WDs and NSs were studied in Ref.~\cite{PC12}.
Under the typical conditions in these layers, the electron
screening proved to increase the enhancement exponent $h$ by
several to tens percent, which translates into a factor of a
few for the reaction rate $\propto\mathrm{e}^h$. The
deviations from the linear-mixing approximation have the opposite effect
of a similar magnitude. Therefore, the corrections
beyond the linear mixing \cite{ChugunovDW09b} must be considered only
together with the electron polarization. In some cases the
two effects nearly compensate each other.

All the discussed corrections, except the quantum one,
proved to have almost no effect on the positions of the
carbon and oxygen ignition curves. In the WDs, the
heat produced by the nuclear reactions is evacuated by
neutrino emission. In this case,  the position of the
ignition curves may be even stronger affected by the current uncertainties
in the neutrino reaction rates in dense plasma environment
than by the departures from the linear-mixing approximation
or by electron-polarization corrections. In
the NSs, the heat is not only taken away by neutrinos, but
also effectively sinks through the envelope. In the latter
case, the one-zone approximation of the heat
diffusion is often applied to the analysis of stability of
the nuclear fusion (e.g.,
Refs.~\cite{BrownBildsten98,Gasques_ea07}).
We have found \cite{PC12} that it is more important to go
beyond the one-zone approximation than to introduce all
other corrections mentioned above. In magnetars (NSs with
superstrong magnetic fields of 
$10^{14}$\,--\,$10^{15}$~G) an
account of the magnetic modification of the heat transport
coefficients is equally significant.

\begin{figure}
\includegraphics[height=72mm]{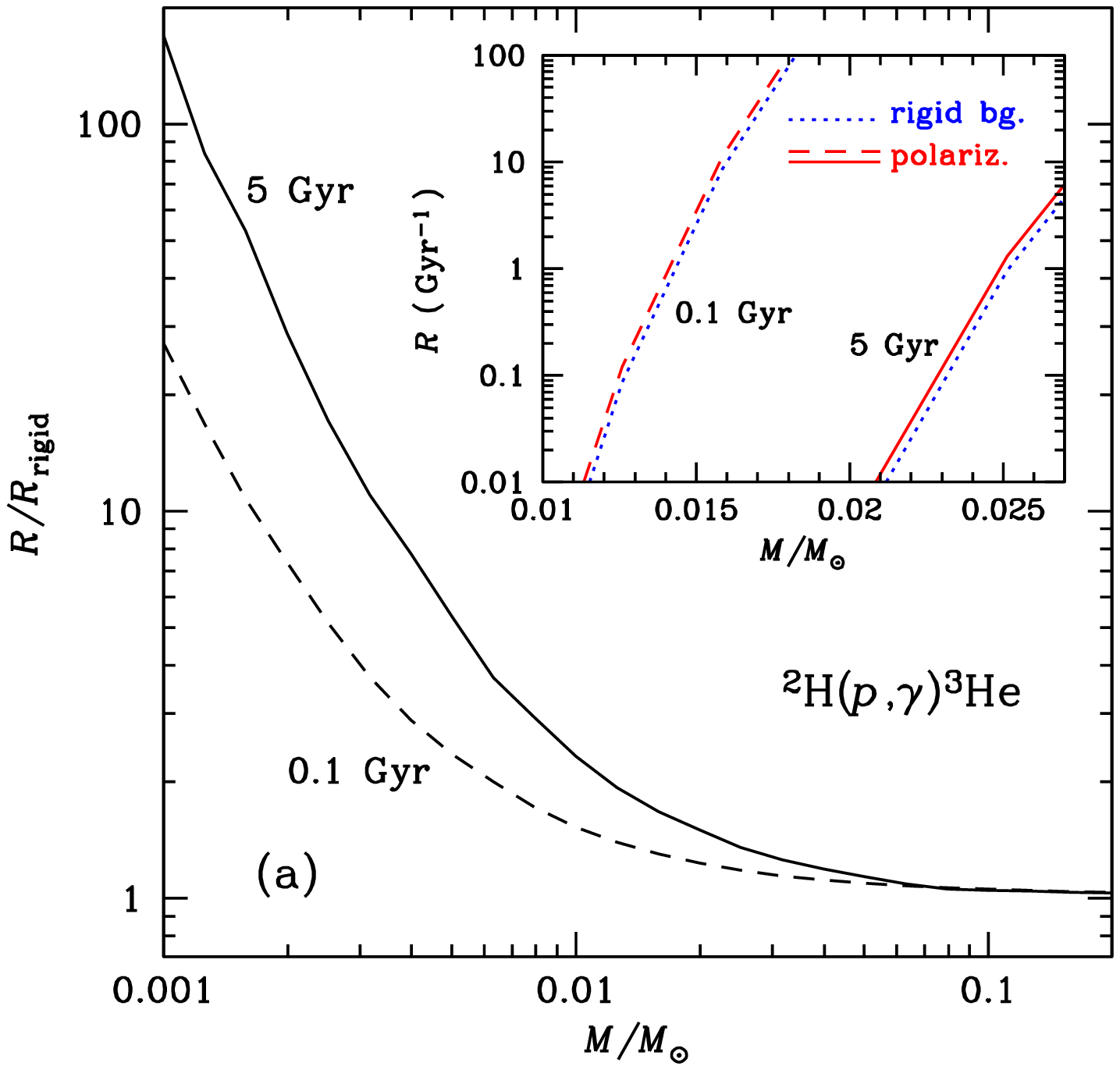}
\hfil
\includegraphics[height=72mm]{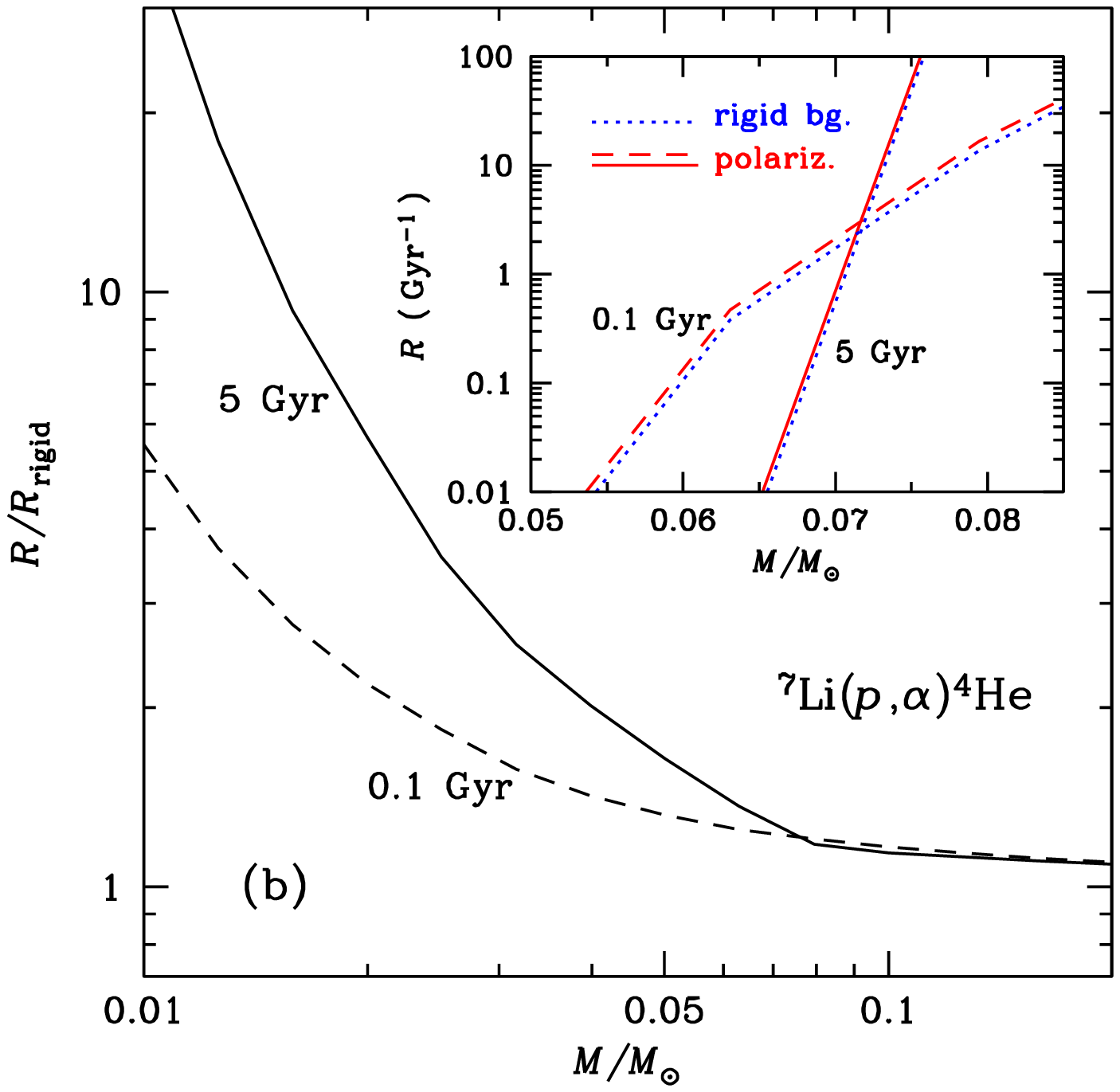}
\caption{(\textbf{a})
Ratios of the deuterium burning rates with account
of electron screening to the rates in the rigid-background
model at the same
temperatures and densities as in Fig.~\ref{fig:enhLMS1}. The
inset shows the absolute reaction rates per one nucleus with
account
of electron screening (solid line for age 5 Gyr, dashed line
for 100 Myr) and for the rigid background (dotted lines).
\textbf{(b)} The same for lithium burning.}
\label{fig:rLMS1}
\end{figure}

\subsection{Low-mass objects}

As another astrophysical example, let us consider the
electron-screening effect on nuclear fusion in low-mass
stars (LMSs) and substellar objects
(SSOs; see Ref.~\cite{CB00} for a review). Important
indicators of the ages and masses of these objects are the so
called lithium and deuterium tests, which are based on
depletion of lithium and deuterium by nuclear burning.

Figure~\ref{fig:enhqHD} displays the enhancement exponent
$h$, normalized with respect to $h_\mathrm{SVH}$,
for the reaction $p+d\to{ }^{3}$He$+\gamma$, 
in different approximations. The accurate result
is compared to the result of application of the
linear-mixing approximation
in the cases of polarizable electron background according to
\req{hLMR} and rigid background
according to \req{hLMRii}, with and without the quantum
corrections.

We note that the simple Salpeter -- Van Horn approximation
(\ref{SVH}) performs  surprisingly well: its accuracy in the
LMS-SSO conditions, as we see in Fig.~\ref{fig:enhqHD}, is
within a few tens percent. We recall that in WD-NS
conditions its accuracy is still better, typically a few
percent \cite{PC12}. The quantum effects in
Fig.~\ref{fig:enhqHD} are significant only at $\rho\gg10^3$
\gcc. From the lower panel of Fig.~\ref{fig:enhLMS1} we see
that such densities are not reached in the LMS-SSO
conditions, therefore the quantum effects do not play role
in theoretical models of nuclear fusion in LMSs and SSOs
(unlike the NS-WD case \cite{PC12}).

Figure~\ref{fig:enhLMS1} shows the dependences of LMS-SSO
central densities and temperatures at two characteristic ages of
these objects (the two lower panels, from Ref.~\cite{CB00}),
together with the respective normalized enhancement
exponents $h/h_\mathrm{SVH}$ with and without the
rigid-background approximation. As previously, we see that
the corrections beyond the Salpeter -- Van Horn
approximation are of the same magnitude for the rigid and
polarizable electron backgrounds (to $\sim30$\% in
this case), but of different sign.

This difference can  translate into factors of a few
 for the reaction rate $R_{12}\propto\mathrm{e}^h$
in the objects of very small mass, because they are relatively
cool and therefore have a large factor $h=H_{12}(0)/T$ in
their central parts.
Figure~\ref{fig:rLMS1}a demonstrates this for the SSO
deuterium fusion. 
For masses $M\sim 10^{-2}M_\odot$, the minimum mass for
D-fusion \cite{CB00},
where $M_\odot$ is the solar mass,
the electron polarization effect changes $R_{12}$ by a factor of
1.5\,--\,2. This change does not significantly affect the
deuterium depletion curves and therefore is unimportant for
the mass-age deuterium test. As seen from the inset in
Fig.~\ref{fig:rLMS1}, the corresponding corrections are
of the order of a few$\times10^{-4}M_\odot$, which is
astrophysically negligible.

Figure \ref{fig:rLMS1}b shows the rates of the reaction
$^7$Li$+p\to2{} {}^4$He. This case is similar to the
previous one, with the difference that the latter reaction
takes place for more massive objects. At very small masses
the difference in the reaction rates calculated with and
without the allowance for the electron polarization is huge,
but astrophysically unimportant, because these rates are so
low that one may neglect this reaction altogether. At
contrast, for $M\approx0.07\,M_\odot$ this reaction
is crucially important for stellar diagnostics, but in this
case the polarization correction is smaller, and it
translates into a negligible correction for the mass-age
relation.

\section{Remarks on some controversial approaches}
\label{sect:deviant}

\subsection{Yukawa potential}

For an arbitrary degree of degeneracy  (but at not too
strong Coulomb coupling; see \cite{Chabrier90}),  the
screened interaction between ions is approximately described
by the Yukawa potential $(Z_1 Z_2 e^2/r) \exp(-r/D)$, where
$D$ is given in \req{DH}. Pollock and Militzer
\cite{PollockMilitzer04} 
studied the contact probabilities of Yukawa systems with
the intention to simulate the electron-screening effect (see
also Ref.~\cite{RosenfeldChabrier89}).
Based on these simulations, they arrived at the conclusion that 
electron screening ``reduces the enhancement
effect,'' in obvious contradiction with our findings
above and with the earlier results
\cite{SalpeterVH69,DeWittGC73,YaSha,SahrlingChabrier98,Kitamura00}.

We note, however, that the Yukawa model corresponds to the
Thomas-Fermi limit, $\epsilon(k)\sim 1+(\kTF/k)^{2}$, for
the static dielectric function $\epsilon(k)$, which is only
justified at $k\ll\kTF$ (see, e.g.,
\cite{GalamHansen76}). Therefore, this model is
inappropriate at short distances (i.e., large wavenumbers
$k$). In particular, it is not applicable for the evaluation
of the screening potential at zero separation, $H_{12}(0)$.
Therefore, a Yukawa system cannot correctly reproduce the
effect of electron polarization on the nuclear fusion rates.
This fact was recognized by Ichimaru
\cite{Ichimaru93}, who mentioned two opposite effects of
electron screening: first, the binary repulsive potentials
between reacting nuclei are reduced by electrons
(``short-range effect''), which increases $H_{12}(0)$;
second, the reduction of particle interactions by the
screening affects the many-body correlation function in such
a way that it decreases $H_{12}(0)$  (``long-range
effect''). In real electron-ion plasmas (without the Yukawa
approximation) the first effect overpowers the second one.
The Yukawa model grasps the second effect, but misses the
first, dominant one.

\subsection{``Quantum tail'' in energy distribution}

Starostin and coworkers
\cite{AleksandrovStarostin98,StarostinSF00,Starostin-ea02,EletskiiST05,Fisch-ea12}
noted that in dense plasmas, in addition to the potential
lowering that results
in the enhancement factor discussed above, there is another
effect capable to modify the reaction rates. Because a state
with definite momentum has a finite lifetime, its momentum
distribution (or, equivalently, the distribution of kinetic
energies) is broadened due to Heisenberg uncertainty
principle. Therefore the probability to find a pair of
nuclei in a state with a high center-of-mass energy is
larger than predicted by the Boltzmann statistics. These
use in \req{R}, instead of
$w_\mathrm{B}(E)$, a modified distribution function 
\beq
w_\mathrm{mod}(E_p) = \int_0^\infty w_\mathrm{B}(E)\,
\phi(E,E_p) \,\dd E,
 \label{w} 
\eeq
 where $\phi(E)$
describes the quantum broadening of the kinetic energy,
defined through the particles' momenta $p$. The authors suggest
to use for this broadening the spectral function
\cite{KadanoffBaym,GalitskiiYakimets}
\beq
 \phi(E,E_p) =
(\gamma/\pi)\,\big/\,\big[ (E - E_p - \Delta)^2 + \gamma^2 \big],
\label{phi} 
\eeq
 where $\gamma=\gamma(p)=\hbar\nucol$ is a collisional
width, $\nucol$ is an effective collision frequency, and
$\Delta$ is the collisional energy shift, which is 
inessential for our discussion and will be suppressed. 
The use of the Lorentz profile (\ref{phi}) in \req{w}
implies the necessary condition $\gamma\ll T$, which is
satisfied in all examples discussed hereafter.
For Coulomb scattering of a light particle with charge $Z_1 e$,
momentum $p$ and velocity $v$ from ambient heavy particles
with charges $Z_2 e$,  the collision frequency is $\nucol = n_2
\langle\sigma_\mathrm{coll}  v\rangle = \big[4\pi n_2 (Z_1
Z_2 e^2)^2 / p^2 v \big] \Lambda(p)$, where
$\sigma_\mathrm{coll}$ is an effective collisional cross
section and $\Lambda(p)$ is a Coulomb logarithm. For
estimates we will use this formula for any
particle masses and adopt the nonmagnetic nonrelativistic limit of the
transport Coulomb logarithm from Ref.~\cite{PY96}: 
$\Lambda(p) = 0.5\,\ln(1+u)-0.5u/(1+u)$,
 where $u=\big[ (D_\mathrm{ion}^2
+ (2a_i/3)^2)^{-1} + \kTF^2 \big]\,(\hbar/2p)^{2}$
(numerically, $\Lambda\sim1$\,--\,10 in the examples below).

In the limit $\gamma\to0$, the function $\phi$ turns into a Dirac delta
function, and the Boltzmann distribution is recovered. 
However, for finite $\gamma$ and for  energies much higher
than the thermal energy, so that 
\beq
   E_p/T \gg \ln(\pi E_p^2/\gamma T),
\label{qtE}
\eeq
 the exponential  decay of $w_\mathrm{B}(E)$ is
overpowered by so called ``quantum tail'' $w_\mathrm{qt}(E_p)
\to \gamma T / \pi E_p^2$. For nonrelativistic Coulomb
particles, $\gamma(p)\propto E_p^{-3/2}$, therefore the tail
decays as $w_\mathrm{qt}\propto E_p^{-7/2}$.\footnote{The authors 
\cite{Fisch-ea12} obtain $w_\mathrm{qt}\propto E_p^{-4}$,
because they replace $v=\sqrt{2E_p/m_{12}}\,$ by
$\sqrt{2E/m_1}$ in $\langle\sigma_\mathrm{coll}v\rangle$, 
which introduces an error in \req{w}
at $E_p\gg T$. However, this difference in
the power index does not qualitatively change any results
or conclusions. We note in passing that the parallel to
Kimball's power law $p^{-8}$ \cite{Kimball}, drawn in
\cite{StarostinSF00,Starostin-ea02,EletskiiST05,Fisch-ea12},
appears ungrounded, because the Kimball's result is specific
to the distribution of fast particles scattered by a
potential with asymptotic behavior $\sim r^{-1}$ at
$r\to0$, whereas the collisional broadening
 is present for any scattering potential \cite{KadanoffBaym}.}

This approach was criticized by Bahcall et
al.~\cite{Bahcall-ea02} and by Zubarev \cite{Zubarev08}.
Bahcall et al.\ merely state that the term
$\dd^3p_1\dd^3p_2$ in the quantum-mechanical expression
$R\propto \int\int \dd^3p_1\dd^3p_2\exp(-E/T)\,|\langle
f|H|i\rangle|^2$ for the reaction rate represents the
density of states and should not be confused with the
expectation values of particle momenta, distributed
according to $w_\mathrm{mod}(E_p)$. While the statement is
certainly true in general, the argument misses the point because it
says nothing about $\langle
f|H|i\rangle$, which needs not be just a
first-order perturbation matrix element. Indeed, the kinetic
Green function derivation of the distribution (\ref{w})
\cite{GalitskiiYakimets,Starostin-ea02} implies that
multiple scattering is taken into account in addition to the
fusion matrix element \textit{per se}. Thus one may say that
$\phi(E)$ is effectively contained in $|\langle
f|H|i\rangle|^2$ in the above expression for $R$.

\begin{figure}[float]
\sidecaption
\includegraphics[height=80mm]{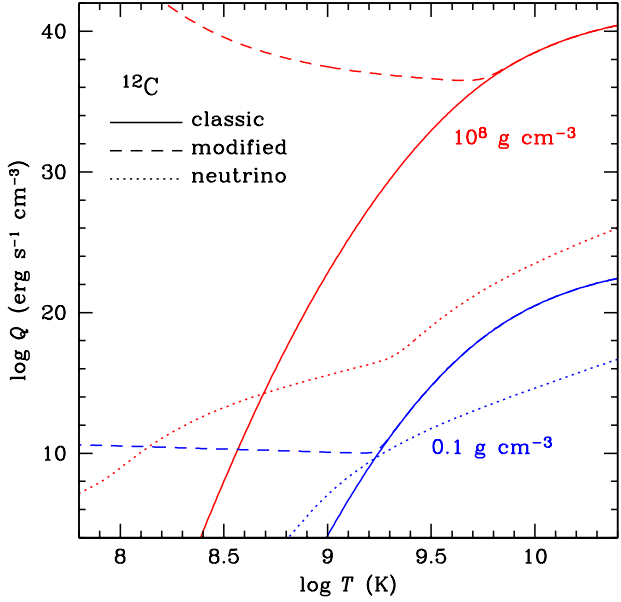}
\caption{Power of the carbon fusion reaction  as function of
temperature $T$ at constant density $\rho$ according to the
conventional calculation (solid lines) and with statistical
averaging over momenta including ``quantum tails'' in their
distribution according to \req{w} (dashed lines). For
comparison, neutrino emission power is shown by dotted
lines. The upper line of each type corresponds to
$\rho=10^8$ \gcc{} and the lower line to $\rho=0.1$ \gcc.}
\label{fig:Qqt}
\end{figure}

Zubarev \cite{Zubarev08} put forward another argument. He
noted that the $pp$-fusion rates calculated with
$w_\mathrm{B}(E)$ and with $w_\mathrm{mod}(E)$ differ by
three orders of magnitude for the Sun, and concluded that in
the second case ``one has neglected the coupling between the
various probability amplitudes of velocity which is
introduced by the quantum uncertainty.'' These statements
are refuted in \cite{Fisch-ea12}, but if there were
indeed such a large difference, the second
(nonstandard) method of averaging must have a flaw.
Indeed, the agreement between the current stellar evolution theory
and observations does not leave room for such a huge change
of the basic reaction rates. 

Let us consider the fusion reaction $^{12}$C+$^{12}$C${}\to{}
{}^{24}$Mg, which
plays a crucial role in the theory of WDs, red giants, and
accreting NSs. We calculate the astrophysical factor $S(E)$
for this reaction using the most recent effective potential
model \cite{Yakovlev-ea10} derived from laboratory results.
In Fig.~\ref{fig:Qqt} we show thermonuclear heat rates
$Q_\mathrm{nuc}$ per unit volume as functions of $T$ at
densities $\rho=10^8$ \gcc{} and 0.1 \gcc, calculated with
using the classical statistical weight $w_\mathrm{B}(E)$ and
the modified weight $w_\mathrm{mod}(E)$. The conventional
results  show steep decrease with decreasing temperature
below $T\lesssim10^9$~K. At contrast, the modified
calculation gives a minimum at $T\gtrsim10^9$~K and then an 
increase of the rate with decreasing $T$. The origin of this
behavior is mathematically obvious. At low $T$, the 
denominator in \req{R} remains determined by the Boltzmann part
of $w(E)$ and provides the normalization $\propto\sqrt{T}$,
while the largest contribution to the numerator  comes from
the integration of $S\mathrm{e}^{-2\pi\eta}$ with the weight
$w\approx w_\mathrm{qt}$ over the energies above the Coulomb
barrier, which is temperature-independent. Thus the ratio
becomes $\propto{T}^{-1/2}$. The latter dependence is well
discerned in Fig.~\ref{fig:Qqt} for $T\lesssim10^9$~K and
$\rho=0.1$ \gcc. At $\rho=10^8$ \gcc, there is a stronger
increase of $Q_\mathrm{nuc}$ with decreasing $T$, caused by
the increase of the factor $\mathrm{e}^h$.

From the physics point of view, this enhancement of nuclear
power at low $T$ is unacceptable. For example, it is
incompatible with the existence of carbon WDs. To show this,
we have additionally plotted in Fig.~\ref{fig:Qqt} neutrino
emission rates $Q_\nu$,  calculated following
Ref.~\cite{Yakovlev-ea01} as the sum of the power carried
away by
neutrino emission
due to annihilation of electron-positron pairs, plasmon
decay, and bremsstrahlung. The intersection 
$Q_\mathrm{nuc}=Q_\nu$ is the \emph{ignition point}, beyond
which nuclear burning becomes unstable.  We see that with
the modified statistical averaging the
intersection is absent, i.e., the burning is always unstable. 
If it were true, all carbon WDs should have
exploded, but they do exist. The modified calculation also
predicts cold fusion at the normal conditions, which does not
happen. 

What is the basic flaw of the ``quantum-tail''
calculation? As can
be seen, for example, from 
Ref.~\cite{Starostin-ea02}, \req{w}
is related to a perturbation correction of the order $\hbar^2$ to
the Maxwellian distribution. Different
forms of this correction can be equivalent to the same order. 
For example, the original Wigner expansion of his
probability function in powers of $\hbar^{2}$ \cite{Wigner32} can be
rewritten in several ways:
$\exp(-E_p/T)\,[1 + \hbar^2 g_2 E_p/T^4 + \ldots]
\sim \exp[-E_p/T +  \hbar^2 g_2 E_p/T^4 + \ldots]
\sim \exp[-E_p/(T + \hbar^2 g_2 / T^3 + \ldots)]$
 (cf.~\cite{LaLi-SP1}, \S\,33).
Here, $g_2$ is a coefficient involving average products of
derivatives of the interaction potential.
These different forms,
however, are not equivalent
if $\hbar^2 g_2 E_p/T^4$ is large, which indicates that
applicability of this correction is restricted to
relatively low energies or high temperatures.
However, the ``quantum-tail'' contribution to the
numerator of \req{R} at low $T$
comes mainly from high energies, whose difference from
$T$ exceeds the collisional width $\gamma$
by many orders of magnitude. Therefore we think that 
such evaluation of \req{R} falls
beyond the applicability range of the distribution
(\ref{w}).

The quantum uncertainty of particles' momenta is conjugate
to the quantum uncertainty of their coordinates. The
thermonuclear fusion enhancement and the cold (pycnonuclear)
fusion due to these quantum uncertainties are well known
(e.g., \cite{SalpeterVH69}). However, these effects are important
only at very high densities (e.g., at $\rho \gg 10^9$ \gcc{}
for the carbon fusion \cite{Yakovlev-ea06}).

\section{Conclusions}
\label{sect:concl}

We have studied the effects of electron screening on
thermonuclear reactions in dense plasmas and compared
different approximations to determine plasma enhancement
factors for the nuclear fusion rates. The electron screening
always increases the enhancement effect. The opposite
conclusion may come from using the Yukawa potential model,
which is inappropriate to calculate the contact
probability for fusing nuclei. The method of taking quantum
uncertainties in particles' momenta by a convolution of
Boltzmann and Lorentz distributions, suggested in some
publications, leads to physically unreasonable results. We
argue that it may be a consequence of
violation of applicability conditions of underlying
theory.

Although the electron polarization correction can increase
a fusion rate by orders of magnitude, we find that it does not
significantly affect theoretical models of WDs, NSs, LMSs,
and SSOs. 

\begin{acknowledgement}
We are grateful to the organizers of the 14th International
Conference on Physics of Nonideal Plasmas for invitation to
present our results. We thank A.\,N.~Starostin for drawing our
attention to the ``quantum tail'' problem.
A.Y.P.~acknowledges useful discussions with G\'erard Massacrier and
Dima Yakovlev.
The work of A.Y.P.\ was partially supported by the Ministry
of Education and Science of the Russian Federation (Agreement
No.\,8409, 2012), the Russian Foundation for Basic Research
(RFBR grant 11-02-00253-a), and the Russian Leading
Scientific Schools program (grant NSh-4035.2012.2).
\end{acknowledgement}

\newcommand{\artref}[4]{#2 \textbf{#3}, #4 (#1)}


\begin{thebibliography}{10}

\bibitem{Gamow28}
G.~Gamow,
\artref{1928}{Z.\ Phys.}{51}{204}

\bibitem{Schatzman48}
E.~Schatzman,
\artref{1948}{J.~Phys.\ Radium}{9}{46}

\bibitem{YaSha}
D.\,G.~Yakovlev and D.\,A.~Shalybkov,
\artref{1989}{Sov.\ Sci.\ Rev., Ser.\ E: Astrophys.\ Space
Phys.}{7}{311}

\bibitem{Ichimaru93}
S.~Ichimaru,
\artref{1993}{Rev.\ Mod.\ Phys.}{65}{255}

\bibitem{Bahcall-ea02}
J.\,N.~Bahcall, L.\,S.~Brown, A.~Gruzinov, and
R.\,F.~Sawyer,
\artref{2002}{Astron.~Astrophys.}{383}{291};
erratum: \artref{2002}{Astron.~Astrophys.}{388}{660}

\bibitem{Salpeter54}
E.\,E.~Salpeter,
\artref{1954}{Australian J.~Phys.}{7}{373}

\bibitem{ItohTI77}
N.~Itoh, H.~Totsuji, and S.~Ichimaru,
\artref{1977}{Astrophys.~J.}{218}{477}

\bibitem{SahrlingChabrier98}
M.~Sahrling and G.~Chabrier,
\artref{1998}{Astrophys.~J.}{493}{879}

\bibitem{Kitamura00}
H.~Kitamura,
\artref{2000}{Astrophys.~J.}{539}{888}

\bibitem{Yakovlev-ea06}
D.\,G.~Yakovlev, L.\,R.~Gasques, A.\,V.~Afanasjev, M.~Beard,
and M.~Wiescher,
\artref{2006}{Phys.~Rev. C}{74}{035803}

\bibitem{ChugunovDW09a}
A.\,I.~Chugunov and H.\,E.~DeWitt,
\artref{2009}{Phys.~Rev. C}{80}{014611}
 
\bibitem{ChugunovDW09b}
A.\,I.~Chugunov and H.\,E.~DeWitt,
\artref{2009}{Contrib.\ Plasma Phys.}{49}{696}

\bibitem{Fowler84}
W.\,A.~Fowler,
\artref{1984}{Rev.~Mod.~Phys.}{56}{149}
 
\bibitem{DeWittGC73}
H.\,E.~DeWitt, H.\,C.~Graboske, and M.\,S.~Cooper,
\artref{1973}{Astrophys.~J.}{181}{439}

\bibitem{Jancovici77}
B.~Jancovici,
\artref{1977}{J.\ Stat.\ Phys.}{17}{357}

\bibitem{RosenfeldChabrier89}
Y.~Rosenfeld and G.~Chabrier,
\artref{1989}{J.~Stat.~Phys.}{89}{283}

\bibitem{IchimaruKitamura96}
S.~Ichimaru and H.~Kitamura,
\artref{1996}{Publ.~Astron.~Soc.~Pacific}{48}{613}

\bibitem{HansenVieillefosse76}
J.\,P.~Hansen and P.~Vieillefosse,
\artref{1976}{Phys.~Rev.~Lett.}{37}{391}

\bibitem{ChabrierAshcroft90}
G.~Chabrier and N.\,W.~Ashcroft,
\artref{1990}{Phys.\ Rev. A}{42}{2284}

\bibitem{CP98}
G.\,Chabrier and A.\,Y.~Potekhin,
\artref{1998}{Phys.~Rev. E}{58}{4941}
 
\bibitem{SalpeterVH69}
E.\,E.~Salpeter and H.\,M.~Van Horn,
\artref{1969}{Astrophys.~J.}{155}{183}

\bibitem{PC10}
A.\,Y.~Potekhin and G.~Chabrier,
\artref{2010}{Contrib.~Plasma Phys.}{50}{82}

\bibitem{Chabrier90}
G.~Chabrier,
\artref{1990}{J.\ Phys.\ (Paris)}{51}{1607}

\bibitem{AlaJanco78}
A.~Alastuey and B.~Jancovici,
\artref{1978}{Astrophys.~J.}{226}{1034}

\bibitem{PollockMilitzer04}
E.\,L.~Pollock and B.~Militzer,
\artref{2004}{Phys.~Rev.~Lett.}{92}{021101}

\bibitem{MilitzerPollock05}
B.~Militzer and E.\,L.~Pollock,
\artref{2005}{Phys.~Rev. B}{71}{134303}

\bibitem{PC12}
A.\,Y.~Potekhin and G.~Chabrier,
\artref{2012}{Astron.~Astrophys.}{538}{A115}

\bibitem{BrownBildsten98}
E.\,F.~Brown and L.~Bildsten,
\artref{1998}{Astrophys.~J.}{496}{915}

\bibitem{Gasques_ea07}
L.\,R.~Gasques, E.\,F.~Brown, A.~Chieffi, et al.,
\artref{2007}{Phys.~Rev. C}{76}{035802}

\bibitem{CB00}
G.~Chabrier and I.~Baraffe,
\artref{2000}{Annu.~Rev.~Astron.~Astrophys.}{38}{337}

\bibitem{GalamHansen76}
S.~Galam and J.\,P~Hansen,
\artref{1976}{Phys.~Rev. A}{14}{816}

\bibitem{AleksandrovStarostin98}
N.\,L.~Aleksandrov and A.\,N.~Starostin,
\artref{1998}{Zh.~Eksp.~Teor.~Fiz.}{113}{1661}
[\artref{1998}{JETP}{86}{903}]

\bibitem{StarostinSF00}
A.\,N.~Starostin, V.\,I.~Savchenko, N.\,J.~Fisch,
\artref{2000}{Phys.~Lett. A}{274}{64}

\bibitem{Starostin-ea02}
A.\,N.~Starostin, A.\,B.~Mironov, N.\,L.~Aleksandrov,
N.\,J.~Fisch, and R.\,M.~Karlsrud,
\artref{2002}{Physica A}{305}{287}

\bibitem{EletskiiST05}
A.\,V.~Eletski\u{\i}, A.\,N.~Starostin, and M.\,D.~Taran,
\artref{2005}{Phys.~Usp.}{48}{281}

\bibitem{Fisch-ea12}
N.\,J.~Fisch, M.\,G.~Gladush, Y.\,V.~Petrushevich,
P.~Quarati, and A.\,N.~Starostin,
\artref{2012}{Eur.~Phys.~J. D}{66}{154}

\bibitem{KadanoffBaym}
L.\,P.~Kadanoff and G.~Baym,
Quantum Statistical Mechanics 
(New York: Benjamin, 1962)

\bibitem{GalitskiiYakimets}
V.\,M.~Galitski\u\i\ and V.\,V.~Yakimets,
\artref{1966}{Zh.~Teor.~Eksp.~Fiz.}{51}{957}
[\artref{1967}{Sov.~Phys. -- JETP}{24}{637}]

\bibitem{PY96}
A.\,Y.~Potekhin and D.\,G.~Yakovlev,
\artref{1996}{Astron.~Astrophys.}{314}{341}

\bibitem{Kimball}
J.\,C.~Kimball,
\artref{1975}{J.~Phys.~A: Math.~Gen.}{8}{1513}

\bibitem{Zubarev08}
A.\,V.~Zubarev,
\artref{2008}{J.~Phys. A: Math.~Theor.}{41}{312004}

\bibitem{Yakovlev-ea10}
D.\,G.~Yakovlev, M.~Beard, L.\,R.~Gasques, and M.~Wiescher,
\artref{2010}{Phys.~Rev. C}{82}{044609}

\bibitem{Yakovlev-ea01}
D.\,G.~Yakovlev, A.\,D.~Kaminker, O.\,Y.~Gnedin, and P.~Haensel,
\artref{2001}{Phys.~Rep.}{354}{1}

\bibitem{Wigner32}
E.~Wigner,
\artref{1932}{Phys.~Rev.}{40}{749}

\bibitem{LaLi-SP1}
L.\,D.~Landau and E.\,M.~Lifshitz,
{Statistical Physics}
(Theoretical Physics, Vol.~5)
(Butterworth-Heinemann, 1980)

\end{thebibliography}
\end{document}